\documentclass[12pt,preprint]{aastex}
\shorttitle{GLOBULAR CLUSTER BLUE STRAGGLERS}
\shortauthors{Piotto et al.}

\begin{document}

\def\hst{{\sl HST}}

\title{Relative Frequencies of Blue Stragglers in Galactic Globular
Clusters:\\Constraints for the Formation Mechanisms\footnote{Based on
observations with the NASA/ESA {\it Hubble Space Telescope}, obtained at
the Space Telescope Science Institute, which is operated by AURA, Inc.,
under NASA contract NAS 5-26555.}}

\author{Giampaolo Piotto\altaffilmark{1},
Francesca De Angeli\altaffilmark{1},
Ivan R.\ King\altaffilmark{2},
S.\ G.\ Djorgovski\altaffilmark{3},
Giuseppe Bono\altaffilmark{4},
Santi Cassisi\altaffilmark{5},
Georges Meylan\altaffilmark{6},
Alejandra Recio-Blanco\altaffilmark{1},
R.\ M.\ Rich\altaffilmark{7},
Melvyn B. Davies\altaffilmark{8}
}

\altaffiltext{1}{Dipartimento di Astronomia, Universit\`a
di Padova, Vicolo dell'Osservatorio 2, I-35122 Padova, Italy;
piotto@pd.astro.it, deangeli@pd.astro.it, recio@pd.astro.it}

\altaffiltext{2}{Department of Astronomy, University of Washington,
Box 351580, Seattle, WA 98195-1580; king@astro.washington.edu}

\altaffiltext{3}{California Institute of Technology, MS 105-24,
Pasadena, CA 91125; george@astro.caltech.edu}

\altaffiltext{4}{Osservatorio Astronomico di Roma, Via Frascati 33,
00040 Monte Porzio Catone, Italy; bono@mporzio.astro.it}

\altaffiltext{5}{Osservatorio Astronomico di Collurania, via M. Maggini,
64100 Teramo; cassisi@astrte.te.astro.it}

\altaffiltext{6}{Space Telescope Science Institute, 3700 San Martin Drive,
Baltimore MD 21218, U.S.A., gmeylan@stsci.edu}

\altaffiltext{7}{Department of Physics and Astronomy, Division of
Astronomy
and Astrophysics, University of California, Los Angeles, CA 90095-1562;
rmr@astro.ucla.edu}

\altaffiltext{8}{Department of Physics and Astronomy, University of
Leicester, Leicester, LE1 7RH, UK}

\begin{abstract}
We discuss the main properties of the Galactic globular cluster (GC)
blue straggler stars (BSS), as inferred from our new catalog containing
nearly 3000 BSS. The catalog has been extracted from the photometrically
homogeneous $V$ vs. ($B-V$) color-magnitude diagrams (CMD) of 56 GCs,
based on WFPC2 images of their central cores.  In our analysis we used
consistent relative distances based on the same photometry and
calibration.  The number of BSS has been normalized to obtain relative
frequencies ($F_{\rm BSS}$) and specific densities ($N_{\rm S}$) using
different stellar populations extracted from the CMD. The cluster
$F_{\rm BSS}$ is significantly smaller than the relative frequency of
field BSS.  We find a significant anti-correlation between the BSS
relative frequency in a cluster and its total absolute luminosity
(mass).  There is no statistically significant trend between the BSS
frequency and the expected collision rate.
$F_{\rm BSS}$ does not depend on other cluster parameters, apart from a
mild dependence on the central density.  PCC clusters act like normal
clusters as far as the BSS frequency is concerned.  We also show that
the BSS luminosity function for the most luminous clusters is
significantly different, with a brighter peak and extending to brighter
luminosities than in the less luminous clusters.  These results imply
that the efficiency of BSS production mechanisms and their relative
importance vary with the cluster mass.
\end{abstract}


\section{Introduction}
Globular Clusters (GCs)
are important astrophysical laboratories for
investigating the stellar dynamics and stellar evolution of low-mass
stars
(e.g., Meylan \& Heggie 1997).  In recent years, it became
clear that we can not study these two astrophysical processes
independently if we want to understand GCs and properly address
several long-standing problems concerning their stellar content.

Among the most puzzling products of the interplay between stellar
evolution and dynamics are the blue straggler stars (BSS). This group of
stars was originally identified by Sandage (1953) in the cluster M3 as a
bluer and brighter extension of the main sequence (MS) turn-off (TO)
stars.  At present, the most popular mechanisms suggested to account for
their origin are {\em primordial binary evolution} (McCrea 1964), i.e.,
mass transfer and/or coalescence in primordial binary systems (Carney et
al.\ 2001), and {\em collisional merging}, i.e., the collision of single
and/or binary systems (Bailyn 1995).  Unfortunately, current photometric
investigations do not allow us to figure out the mechanism that triggers
the formation of BSS in GCs, and indeed it has been suggested that both
primordial binary evolution and collisions are probably at work in
different clusters (Ferraro, Fusi Pecci, \& Bellazzini 1995; Piotto et
al.\ 1999, Ferraro et al. 2003), or even within the same cluster
(Ferraro et al.\ 1997).  The observational scenario concerning BSS
formation has been recently enriched by the results of a spectroscopic
survey by Preston \& Sneden (2000, hereafter PS00). On the basis of
multi-epoch radial velocity data of field blue metal-poor (BMP) stars,
PS00 found that more than 60\% of the stars in their sample are
binaries.  On the basis of empirical evidence, PS00 concluded that at
least 50\% of BMP stars are BSS. Moreover, PS00 suggested that the BSS
in their sample must have formed via mass transfer in binaries.
Finally, PS00 found that the specific frequency of BSS in the local halo
is an order of magnitude larger than in GCs.  This discrepancy 
opens several new questions concerning the origin of
field and cluster BSS.

In an attempt to better understand the properties of BSS stars in GCs,
we took advantage of our homogeneous database of color-magnitude
diagrams (CMD) from WFPC2 images (Piotto et al.\ 2002) to select
a sample of nearly 3000
BSS in 56 GCs characterized by different morphological and
dynamical properties. 
In
this paper we exploit the new BSS catalog to investigate
empirically whether the BSS population is related to any of the
properties of the parent GC. Here we present the results we
believe to be the most relevant and original.  The entire BSS photometric
catalog and further details on the BSS extraction 
will be published in
a forthcoming paper (De Angeli et al.\ 2004), and it
will become available at the Padova Globular Cluster Group web site.

\section{THE DATABASE OF BSS IN GCs}
\label{database}
We have recently completed our HST/WFPC2 snapshot project (GO 7470, GO
8118, GO 8723). By adding the data from our former GO 6095, and
similar data from the archive (i.e., WFPC2 images collected in the
cluster center, in the F439W and F555W bands), we have obtained a total of
74 CMDs. They can be found on the Padova Globular Cluster Group web
pages (http://dipastro.pd.astro.it/globulars).  Details on the data
reduction are in Piotto et al.\ (2002). Here suffice it to say that all
the data have been processed in the same way: instrumental photometry
with DAOPHOT/ALLFRAME, CTE correction and calibration
to the $B$ and $V$ standard systems following Dolphin
(2000). 
Artificial star experiments have been run for all the clusters to
estimate the completeness of star counts. Completeness corrections have
been always applied when necessary. For the BSS sample the completeness
was typically larger than 90\%.

BSS are present in all 74 GCs of our catalog.  However, 18 CMDs were
too contaminated by background/foreground stars, or heavily affected by
differential reddening, to allow us a reliable selection of BSS.
The final catalog,
includes 2798 BSS 
candidates in 56 GCs (actual stars, before completeness correction), i.e., about
five times the number published in previous catalogs (Fusi
Pecci et al.\ 1993; Sarajedini 1993).

The BSS in our sample share one common feature: in the inner region
mapped by our WFPC2 images, the radial distribution of BSS is more
centrally peaked than that of any other cluster population. In fact, we
performed several Kolmogorov-Smirnov tests on the radial
distributions of HB, RGB, and BSS, and we found that the BSS are
more centrally concentrated at the level of $99.9$\%, apart
from a few cases where the small number of stars prevents any
statistically significant test.

\section{THE RELATIVE FREQUENCY OF BSS}
\label{freq}

In this section, we investigate whether the number of BSS is correlated
with any of the physical and morphological parameters of their parent
GCs. In order to compare the number of BSS in different GCs properly, we
must adjust it to allow for how much of each cluster we sampled.  To
this end, a number of different specific frequencies (defined as the
ratio between the number of BSS and a reference population) have been
used in the literature (cf. Ferraro et al.\ 1995).
We estimated the specific frequencies by normalizing the number of
BSS to the HB ($F^{\rm HB}_{\rm BSS}$) or the RGB ($F^{\rm RGB}_{\rm
BSS}$) stars.  Interestingly, the results discussed below do not depend
on which specific frequency we choose.  Therefore, in the following we
will adopt $F^{\rm HB}_{\rm BSS}$, and will refer to it as $F_{\rm
BSS}$.  The numbers of BSS and HB stars have been corrected for
completeness 
(details in Piotto et al.\ 2002) 
before calculating
$F_{\rm BSS}$.

\begin{figure}
\plotone{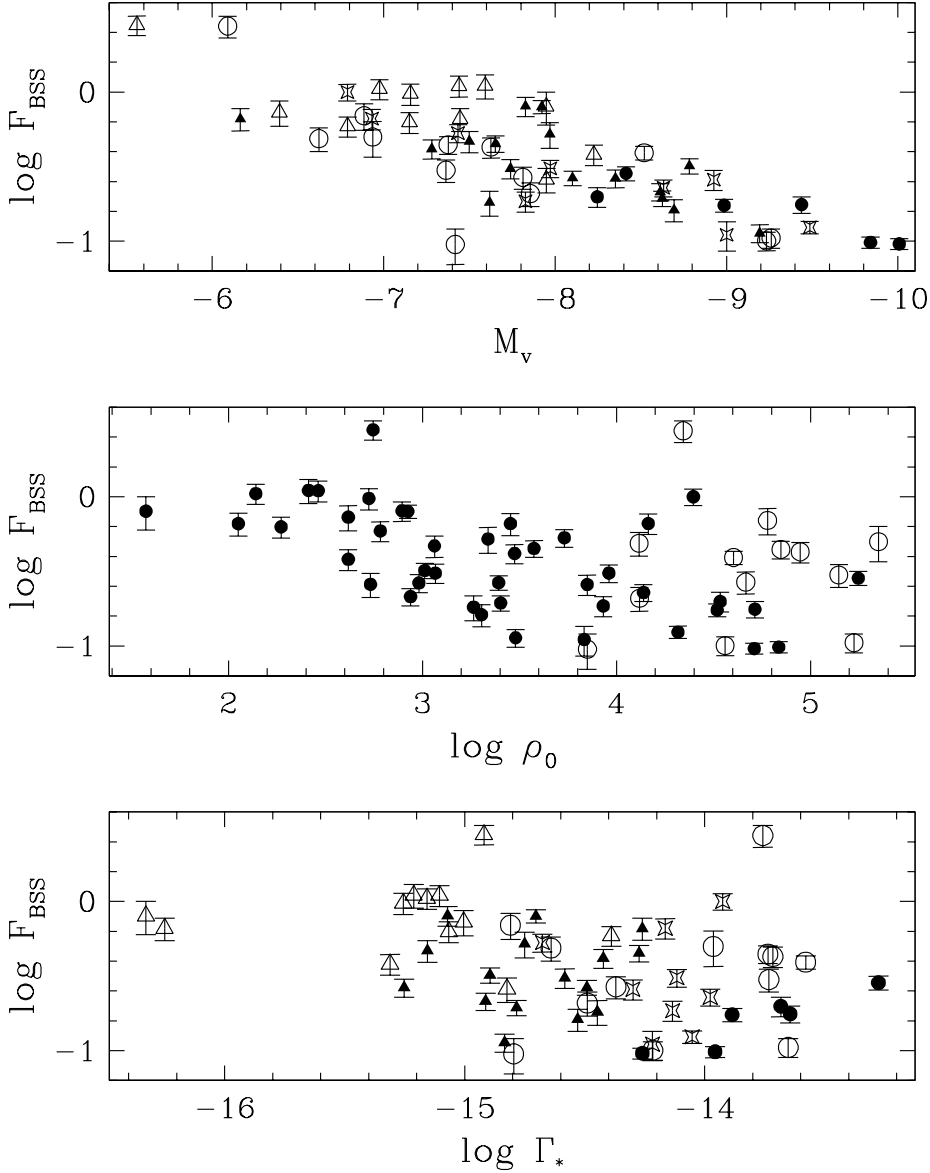}
\figcaption{BSS relative frequency as a function of the integrated absolute
magnitude of the cluster ({\it top panel}), the central density
({\it middle panel}), and of the
collision rate ({\it bottom panel}).
Different symbols are used (in the top and bottom panels) for
clusters with different central densities:
$\log\rho_0<2.8:$ {\it open triangles};
$2.8<\log\rho_0<3.6:$ {\it filled triangles};
$3.6<\log\rho_0<4.4:$ {\it crosses};
$\log\rho_0>4.4:$ {\it filled circles}.
In all panels, PCC clusters are open circles.
}
\end{figure}

The top panel of Fig.\ 1 shows $F_{\rm BSS}$ as a function of
the integrated visual absolute magnitude of the cluster,
from the integrated visual apparent magnitudes and the
reddening values in the Harris catalog and the apparent distance
moduli derived by Recio-Blanco et al.\ (2004a) for the same
clusters by following the procedure outlined by Zoccali et al.\ (2000).
Even though the distribution of empirical data still shows a
large scatter for $-8 \le M_V \le -7.4$, the data in this panel
clearly show a correlation between $F_{\rm BSS}$ and the integrated
absolute
magnitude $M_V$. In particular, the faintest clusters in our sample,
namely NGC 6717 and NGC 6838, present a BSS specific frequency that is
more than a factor of 20 larger than that for the brightest clusters.
This result is in agreement with that found by PS00 on the basis of data
collected from the literature for 30 GCs 
though their plot tends to
flatten for $M_V<-7$, while Fig.\ 1 shows that $F_{\rm BSS}$ decreases
continuously with increasing cluster total luminosity, up to $M_V<-9$.
Interestingly enough, PCC clusters (open circles), with the
exception of NGC 5946 that shows a small value of $F_{\rm BSS}$, 
behave
as normal clusters.

The middle panel of Fig.\ 1 shows that $F_{\rm BSS}$ also depends on
the cluster central density $\rho_0$ (in $L_\odot/pc^3$), though this
correlation is statistically somewhat less significant than the
previous one.  (For internal consistency, we have re-calculated the
central densities using the equations suggested by Djorgovski (1993),
but adopting our new distance moduli and the central surface
brightness in the Harris catalog.)  The $F_{\rm BSS}$ for clusters
with $\log\rho_0>3.2$ shows a large dispersion, and no correlation.
For $\log\rho_0<3.2$, the BSS frequency 
increases with decreasing central density.
Once again, the PCC clusters do not
show any peculiar trend.
We have
also compared $F_{\rm BSS}$ with the concentration parameter $c$ and the
half mass relaxation time $t_{\rm h}$. Here too there is no clear correlation,
though GCs with $\log t_{\rm h}<9$ have, on average, a $F_{\rm BSS}$
three times larger than clusters with a longer relaxation time.

In view of the proposed formation mechanisms for BSS, it is interesting
to check whether 
the $F_{\rm BSS}$
depends on the expected
frequency of stellar collisions. King (2002) demonstrated that the rate
of stellar collisions (per cluster and per year) in a King model GC is
about $\Gamma_{\rm c} = 5\times10^{-15}(\Sigma_0^3r_{\rm c})^{1/2}$,
where $\Sigma_0$ is the central surface brightness in units of $L_{\odot
V}$ pc$^{-2}$ (equivalent to $\mu_v=26.41$), and $r_{\rm c}$ is the core
radius in parsecs (taken from Harris 1996).  In order to calculate the
probability $\Gamma_\star$ that a given star will have a collision in
one year we have divided this collision rate by the total
number of stars ($N_{\rm star}$) in the cluster. This has been estimated
by using the integrated visual absolute magnitude of the cluster,
assuming $M/L=2$ and a typical mass for the colliding stars of $0.4
m_\odot$.  The bottom panel of Figure 1 plots $F_{\rm BSS}$ as a
function of the resulting $\Gamma_\star$. There is no statistically
significant correlation, though we note that, on average, the 
11
clusters with the smallest collision probability
($\Gamma_\star<10^{-15}$) have a BSS frequency 2--3 times higher than
clusters with higher collision rates.  It must be noted here that,
according to the results by PS00, the BSS frequency
of the GCs with the smallest collision probability in our sample is
about 5 times smaller than the BSS frequency in the field, where
collisions are so much rarer.
We note that in the case of PCC clusters, the current values
$\Gamma_\star$ and $\rho_0$ may not be representative of the average
dynamical environment in which currently--observed blue stragglers
have formed. However, Fig\ 1 shows that PCC clusters have BSS
frequencies comparable to normal King--model clusters, possibly
indicating that their dynamical evolution has not affected in a
significant way the BSS formation.

The anti-correlation between $F_{\rm BSS}$ and total cluster
luminosity, the lack of a statistically significant correlation
with the collisional parameter, and the apparently higher relative
frequency of BSS where collision rates are very small are the
most interesting results extracted from our catalog. These empirical
facts are somehow puzzling. In fact, we would have expected more BSS in
clusters where the probability of collision is higher. We will discuss
these results further in the next Section.

The error bars plotted in Fig.\ 1 account for Poisson sampling errors
and the uncertainty in the completeness corrections.  Even if we assume
an upper limit of 0.2 magnitude for the uncertainties in the individual
distance moduli, the correlation shown in Fig.\ 1 represents a robust
empirical result, and the reasons are manifold.  Unlike other data
available in the literature, i) our dataset is photometrically
homogeneous; ii) the star counts included the 2 innermost
arcmin; iii) the star counts of BSS and of the reference populations
have been corrected for incompleteness. As a whole, the present dataset
is only very marginally affected by the thorny statistical problems
affecting previous estimates of BSS specific frequencies (cf.\
discussion in Ferraro et al.\ 1995).

\section{DISCUSSION}
\label{conc}

The statistically significant anticorrelation of the BSS relative
frequency with the integrated luminosity and the 
independence of
the expected collision rate discussed in the previous Section are
noteworthy, and we will concentrate on them.  These observational facts
are complemented by the finding by PS00 that field
BSS have a frequency $F_{\rm BSS}=4.0$,
an order of magnitude larger than the BSS
frequency of the bulk of the GCs.

\begin{figure}
\plotone{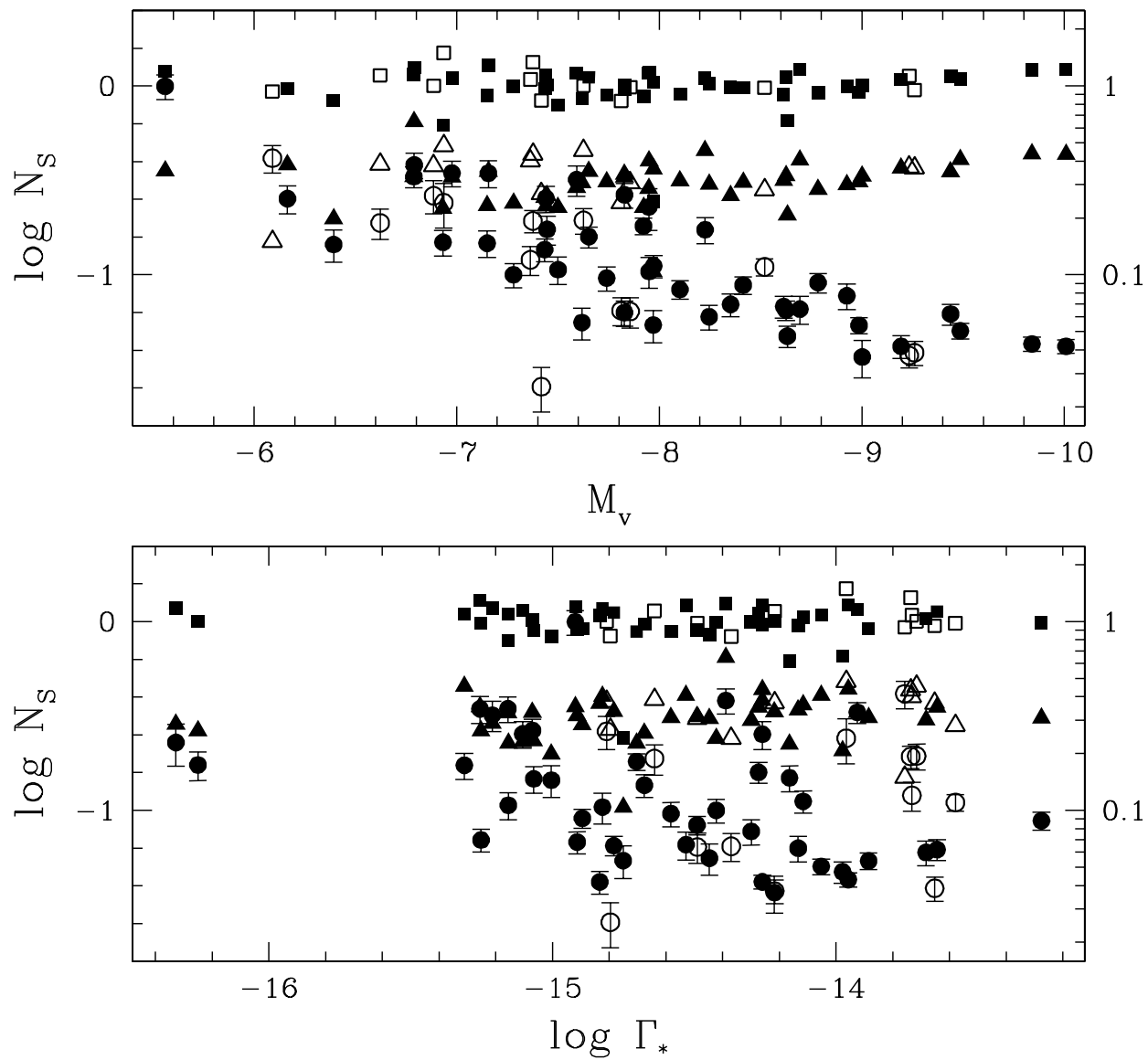}
\figcaption{Number of BSS (circles), HB (triangles), and RGB (squares)
stars per absolute visual flux unit
as a function of the integrated cluster magnitude ({\it top panel}) and
of the collision rate ({\it lower panel}). Open circles represent PCC clusters.}
\end{figure}

Figure 2 shows the same results of Fig.\ 1, in a different way that
may be more enlightening.  Here we look at the number of BSS, HB, and
RGB stars relative to the total flux in the same region.  We call this
quantity $N_{\rm S}$; it is defined by
$$\log N_{\rm S}=\log\left[{N \over (F_{\rm HST}/10^{-0.4 V_{\rm tot}})}
    \cdot {1 \over 10^{-0.4 M_V}}\right],$$
where $N$ is the total number of BSS (or HB or RGB) stars, corrected for
completeness, $F_{\rm HST}$ is the total flux from all the stars that we
measured in the region, $V_{\rm tot}$ is the integrated apparent
magnitude of the cluster, and $M_V$ is its integrated absolute
magnitude.  (Note that our CMDs typically extend well below the turnoff,
so that the contribution from the fainter stars is negligible.)

The first factor in the brackets can be understood as follows: the
quantity $F_{\rm HST}/10^{-0.4 V_{\rm tot}}$ is the fraction of the
total cluster flux that is sampled by the HST field.  If the BSS were
distributed like the flux, then
$N / (F_{\rm HST}/10^{-0.4 V_{\rm tot}})$
would be the number of BSS we would expect if we could observe the whole
cluster.  
In fact, as the BSS are more concentrated to the center than is the
flux, the quantity above is still a reasonable approximation to the
total number of BSS to be expected.  The defect in the approximation
increases the scatter in $N_{\rm s}$, but it does not introduce any
systematic effects, since (as we have verified) there is no
correlation between the fraction of flux included and $M_V$.
The second factor in the brackets is just our previous normalization
to the size of the cluster, but now in luminosity units.

Interestingly enough, Fig.\ 2 confirms that the HB and RGB stars are
very good tracers of the cluster population, as their absolute density
remains constant over more than 4 magnitudes in cluster total
luminosity. This fact removes the risk that the results of Fig.\ 1 might
be due to some anomalous gradient in the distribution of HB and RGB
stars (cf.\ Djorgovski, Piotto \& Capaccioli 1993). Figure\ 2 confirms
that the density of BSS decreases 
with increasing total cluster mass
and
that there is no correlation between the density of BSS
and the collisional parameter. However, we note that, given the small
size of the error bars, the dispersion of the the BSS density is much
larger than the dispersion of the HB and RGB star densities, and that,
as noticed in Fig.~1, clusters with $\Gamma_\star<10^{-15}$ 
have a 2--3 times larger BSS density than clusters
with higher collision rates.
The lack of an overall dependence of $F_{\rm BSS}$ and $N_s$ on the
collisional parameter seems to suggest that 
direct collisions of
single or binary stars 
are
not the main formation mechanism of BSS.
At first glance, the evolution of primordial binaries also does not seem
to be the dominant formation mechanism for BSS in all GCs. In the
simple hypothesis that the binary fraction is the same in all
clusters, we would expect the BSS density to show a behavior similar to
that of the HB and RGB stars, in Fig.\ 2.  On the other hand, the
evolution of primordial binaries is affected by the cluster
environment, and, in particular, it is accelerated in clusters where
the encounter probability is higher.  Indeed, in a paper parallel to
this one, using the mechanism proposed by Davies \& Hansen (1998)
to explain the production of millisecond pulsars in GCs, Davies,
Piotto, \& De Angeli (2004) demonstrate that in clusters with high
encounter probability the formation of BSS from primordial binaries
has been favored in the past.  Now these binaries cannot form BSS
anymore (they have already evolved), and this explains the observed
relative absence of BSS in many high mass, high collision rate
clusters.  It also explains the relatively larger fraction of BSS
among the field stars, where the even lower-density environment makes
the evolution of binaries via encounters slower than in any GC,
allowing them to produce BSS for a more extended time interval
(till the present).

Davies et al.\ (2004) show also that only in the most luminous GCs
(specifically, clusters with $M_V<-8.8$) do the BSS start to be produced
predominantly by stellar collisions.  A better way to characterize the
physical properties of the BSS is to look at their luminosity function
(LF). In order to overcome possible dependencies of the LF on the
cluster metallicity, distance, and reddening, we have divided the
luminosity of each BSS by the turn-off luminosity of the parent cluster.
Figure 3 shows the LFs for GCs with different total luminosity.
The cut in $M_V$ has been set at $M_V=-8.8$, where the theory (Davies et
al.\ 2004) predicts that the BSS should become predominantly
collisional.

\begin{figure}
\plotone{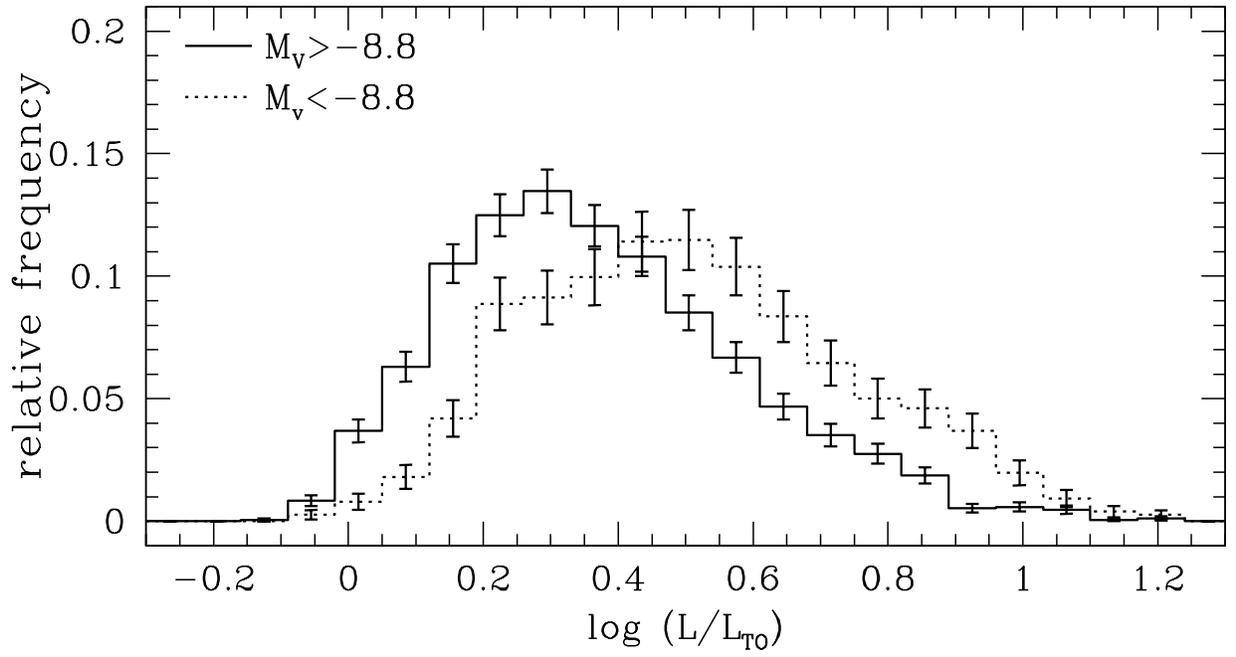}
\figcaption{BSS LFs for clusters with different integrated magnitude.}
\end{figure}

Interestingly enough, clusters with $M_V<-8,8$ have a BSS LF which is
significantly different from the BSS LF of less luminous clusters
(Fig.~3), in that the LFs for the most luminous clusters have a
brighter peak and are significantly shifted toward brighter
magnitudes.  If the relative importance of the BSS production
mechanisms depends on the cluster mass, we would then expect to see a
dependence of the BSS LF on $M_V$, as is observed in Fig.\ 3. In
general, a BSS produced by collision is expected to have a different
luminosity with respect to a BSS from mass transfer or merger of
binaries, due to the resulting interior chemical profile. How much
different is still controversial.  Indeed, recent detailed smoothed
particle hydrodynamic simulations performed by Sills et al.\ (2002)
have shown that collision products are not chemically homogeneous.
This has the effect of producing a BSS structure less blue and less
bright than expexted on the basis of the \lq{fully mixed}\rq\ models
(e.g., Bailyn \& Pinsonneault 1995). Nevertheless, Sills et al.\ (2001)
have also shown that collision products emerge as rapidly rotating
blue stragglers, and so far we lack a full understanding
of the changes in the evolutionary properties due to rotationally
induced mixing.

It is also worth noting that PCC clusters seem to have normal BSS
population. This might be due to the fact that the core collapse
phase is very short, and confined to the very central part of the
clusters, and therefore does not affect the BSS production over
the last few Gyr.

\acknowledgments This research was supported by the Ministero
dell'Istruzione, Universit\`a e Ricerca (PRIN 2001 and PRIN 2002), and
by the Agenzia Spaziale Italiana.  I.R.K.\ and S.G.D.\ acknowledge the
support of STScI Grants GO-6095, 7470, 8118, and 8723.


\end{document}